\documentclass[conference]{IEEEtran}

\IEEEoverridecommandlockouts

\usepackage{cite}

\usepackage[english]{babel}

\ifCLASSINFOpdf
   \usepackage[pdftex]{graphicx}
\else
   \usepackage[pdftex]{graphicx}
   \DeclareGraphicsExtensions{.eps}
\fi

\usepackage{grffile}
\usepackage{epstopdf}

\usepackage[tight,footnotesize]{subfigure}

\usepackage{fixltx2e}

\usepackage{slashbox}

\usepackage[cmex10]{amsmath}
\usepackage{color}
\usepackage{array}
\begin{document}

\title{On the genetic optimization of APSK constellations for satellite broadcasting}

\author{\IEEEauthorblockN{Alessio Meloni\thanks{\copyright 2014 IEEE. The IEEE copyright notice applies. DOI: 10.1109/BMSB.2014.6873465} and Maurizio Murroni}

\IEEEauthorblockA{DIEE - Department of Electrical and Electronic Engineering\\
University of Cagliari\\
Piazza D'Armi, 09123 Cagliari, Italy\\
Email: \{alessio.meloni\}\{murroni\}@diee.unica.it}
}

\maketitle

\begin{abstract}
Both satellite transmissions and DVB applications over satellite present peculiar characteristics that could be taken into consideration in order to further exploit the optimality of the transmission. In this paper, starting from the state-of-the-art, the optimization of the APSK constellation through asymmetric symbols arrangement is investigated for its use in satellite communications. In particular, the optimization problem is tackled by means of Genetic Algorithms that have already been demonstrated to work nicely with complex non-linear optimization problems like the one presented hereinafter. This work aims at studying the various parameters involved in the optimization routine in order to establish those that best fit this case, thus further enhancing the constellation.\\  
\end{abstract}

\begin{IEEEkeywords}
Transmission, Channel coding, modulation, multiplexing, Signal processing for transmission\\
\end{IEEEkeywords}

\IEEEpeerreviewmaketitle

\section{Introduction}

APSK modulation with pre- and post- compensation schemes \cite{gaudenzi2006} is deployed in DVB over satellite standards \cite{DVB2} \cite{meloni2014} for its power and spectral efficiency over nonlinear satellite channels. Nevertheless, for multimedia broadcasting applications, further improvements by means of non-uniform constellations could be obtained. As a matter of fact, multimedia streams employed in digital broadcasting are hierarchical by nature, so that bits associated with transmitted symbols present different error sensitivities. In particular, Most Significant Bits (MSB) affect the transmission more than errors on the Least Significant Bits (LSB). 

Channel coding techniques with Unequal Error Protection (UEP) have been studied in \cite{zoellner2013} for QAM and in \cite{kayhan2012} for APSK, even though this introduces overhead and reduces bandwidth efficiency, which is a critical issue for satellite applications. In \cite{bruggen2005} Modulation with Unequal Power Allocation (MUPA) was proposed as a mean to improve the performance of conventional modulation schemes in case of digital wireless communication systems (e.g. DECT, Bluetooth) which do not include channel coding for some reason thus saving bandwidth. MUPA distributes the available budget power over the QAM symbols according to their sensitivity to channel errors, whereas the average transmission power per symbol remains unchanged. The resulting quality on received data is improved without any increase of transmission bandwidth.
 
The same concept was extended in \cite{angioi2010} to the case of APSK modulations in order to achieve UEP through asymmetric layout of the constellation symbols. The approach used is similar to MUPA in the selection of the opportune radius of the constellation circles and the phase of each symbol. The best numerical solution was obtained solving the optimization problem (OP) of minimizing inter-symbol distortion by means of Genetic Algorithms (GA) \cite{whitley1994}, a numerical search technique used in many fields to solve complex problems which do not allow analytical derivation \cite{lixia2011} \cite{sanna2010}. This work focuses on the OP through careful selection of the parameters involved in GA in order to further improve the constellation performance with respect to \cite{angioi2010}. Moreover, the possibility to drop various symmetric bit allocation constraints is taken into consideration as a mean to further boost the constellation optimality. 

\section{System Model}

First of all let us introduce the model used for the optimization by schematically describing the various building blocks of the communication system.

\begin{figure}[htbp!]
\centering
\includegraphics [width=1 \columnwidth] {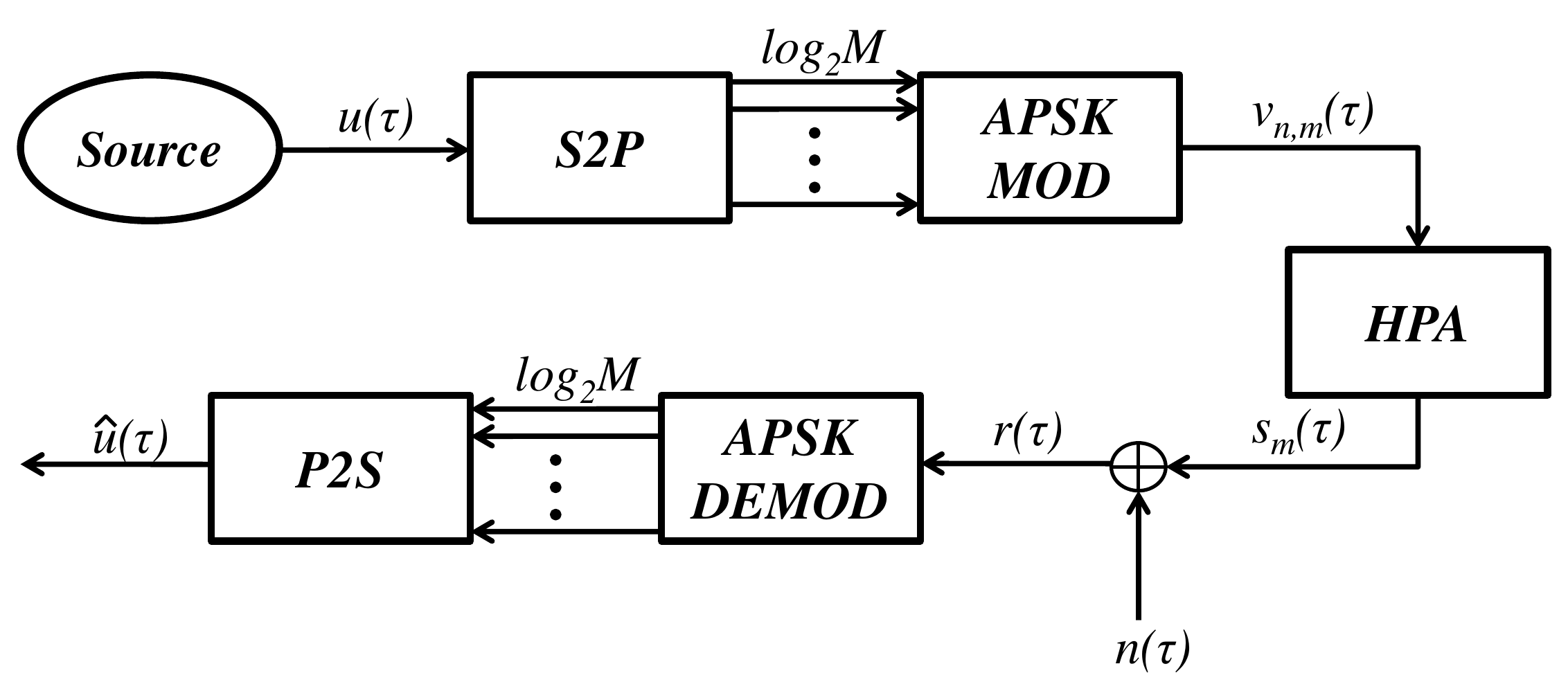}
\caption{System model}
\label{syst_mod}
\end{figure}

\begin{figure}[htbp!]
\centering
\includegraphics [width=1 \columnwidth] {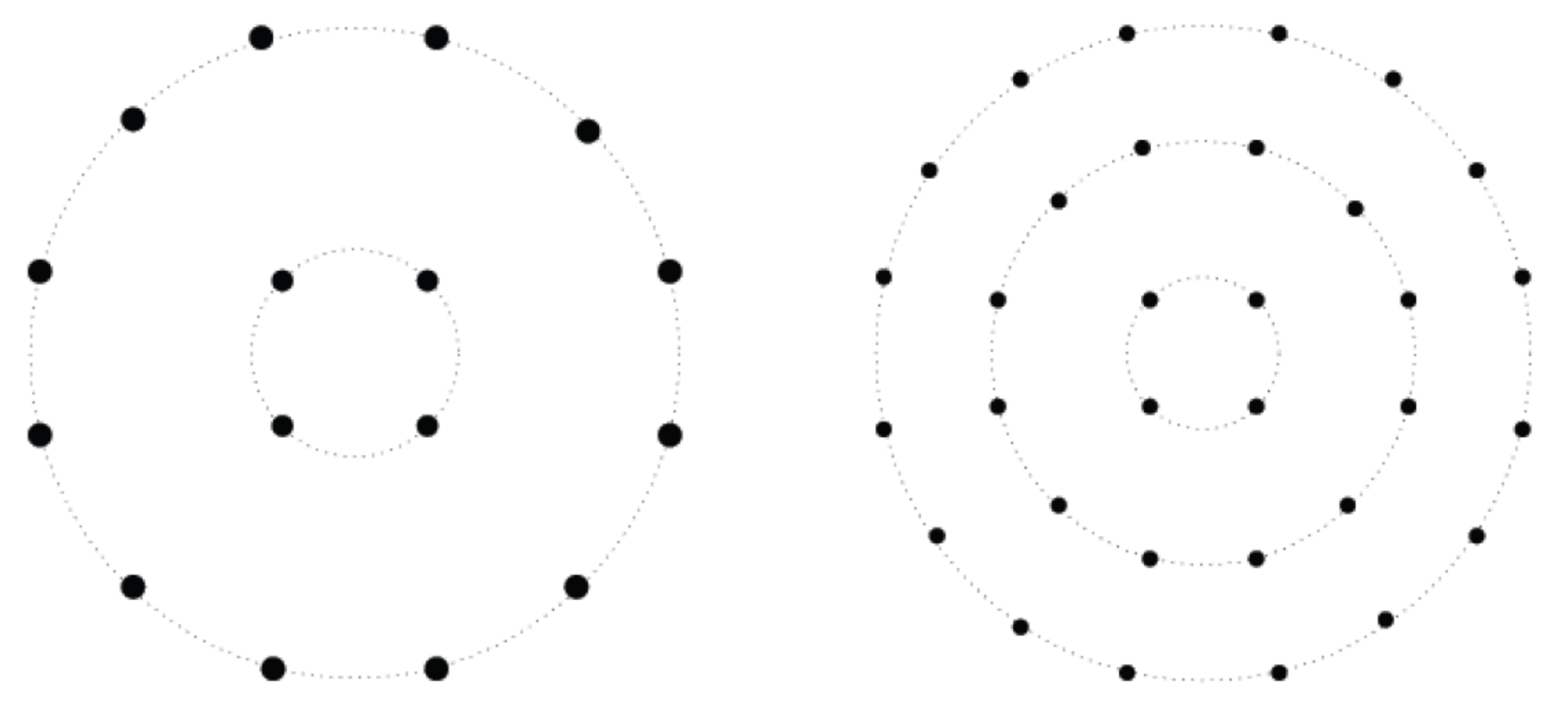}
\caption{16-APSK constellation (left) and 32-APSK constellation (right)}
\label{1632const}
\end{figure}

Considering $M$ to be the alphabet size of the constellation, each symbol will represent a stream $u(\tau)$ of $log_2M$ bits generated by a memoryless source and put in parallel by the serial-to-parallel (S2P) block. These symbols are then modulated by the APSK modulation block giving place to a complex number $v_{n,m}(\tau)=\rho_n\cdot e^{j\theta_m}$ that represents constellation symbols and where $\rho_n$ is the radius of the $n-th$ circle of the constellation and $\theta_m$ is the phase of the $m-th$ symbol. As already claimed in \cite{angioi2010} the distribution of the symbols of the constellation on the various radii is basically free. However, considering the non-linear distorsion introduced by the HPA the best performance is obtained when $4$ symbols are put in the inner circle for 16-APSK (i.e. $4+12$) and $4$ in the inner and $12$ in the medium circle are put for 32-APSK (i.e. $4+12+16$) as shown in Figure \ref{1632const}. 

Once constellation symbols are modulated, they are amplified by the HPA prior to transmission, thus being subject to the non-linear behaviour of the amplifier, whose effects can be modeled using the Saleh model \cite{saleh1981} resulting in the output $s_m(\tau)$. This model distinguishes two effects:
\begin{itemize}
\item the AM/AM non-linear effect that models amplitude distorsions on the input signal;
\item the AM/PM non-linear effect that models phase distorsions on the input signal.
\end{itemize}

In this paper, optimization has been accomplished without taking into consideration the AM/PM non-linear effect due to the HPA (that is known to change the relative position of symbols) since it has been already shown in \cite{gaudenzi2006} that this effect can be easily compensated. Therefore only the AM/AM non-linear distorsion is taken into account through the formula

\begin{equation}
A(\rho)=\frac{a\rho_n}{1+b\rho_n^2}
\end{equation}

where $a=2.1587$ and $b=1.1517$ are standard values of the constants gathered from the literature and obtained by means of curve-fitting techniques. 

\begin{figure}[htbp!]
\centering
\includegraphics [width=0.8 \columnwidth] {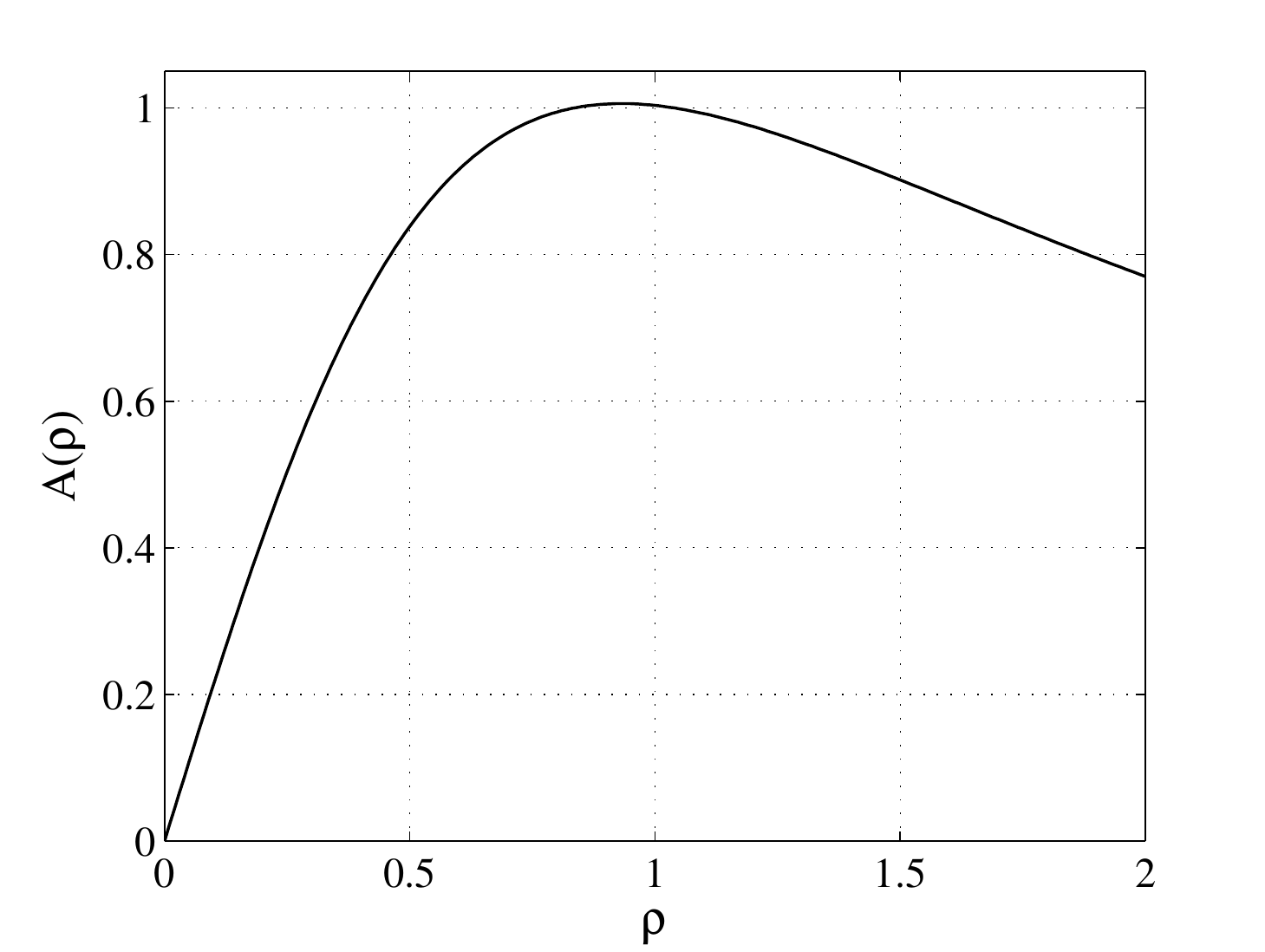}
\caption{AM/AM characteristics \cite{gaudenzi2006}}
\label{saleh}
\end{figure}

Considering the channel to be of the Additive White Gaussian Noise (AWGN) type, transmitted symbols are affected by the addition of a random nuisance signal with zero mean and variance $N_0/2$. Therefore, the received signal at the destination is $r(\tau)=s(\tau)+n(\tau)$. This signal is passed to the demodulator that applies the Maximum Likelihood (ML) criterion in order to estimate the received symbol and it is then converted back from parallel to the serial signal $\hat{u}(\tau)$. The difference between the stream generated by the memoryless source and the one estimated at the receiver can be computed as 
\begin{equation}
d(\tau)=[u(\tau)-\hat{u}(\tau)]
\end{equation}
 and it is called distorsion. The aim of the GA presented in the next section is to minimize this distorsion. Therefore the minimum square error 
\begin{equation}
MSE=E\{[d(\tau)]^2\} 
\end{equation}
represents our optimization criterion.    

\section{Genetic Algorithms optimization}\label{GA}

At each iteration $n$, a GA gives birth to a generation $G_n$ of potential solution vectors (also called chromosomes $\gamma_i$) that constitute the population of size $p$ of the OP:
\begin{equation}
G_n=\{ \gamma_1^{(n)}, \gamma_2^{(n)},..., \gamma_i^{(n)},..., \gamma_{p}^{(n)} \}
\end{equation}

 A vector of fitness scores $S_n$ is also calculated for each generation using the objective function $R$:
\begin{equation}
S_n=\{ R(\gamma_1^{(n)}), R(\gamma_2^{(n)}),..., R(\gamma_i^{(n)}),..., R(\gamma_{p}^{(n)}) \}
\end{equation}
The chromosomes with the highest fitness score are meant to be the closest to the desired solution and are thus selected for surviving and giving place to the next generation.

The next generation is created in three steps:
\begin{enumerate}
\item \textit{Selection} of the part of population with the best fitness score that will be parents for the next generation;
\item \textit{Crossover} of selected parents according to a mixing criterion in order to give birth to a number of children from each couple that will constitute the next generation;
\item \textit{Mutation} of a percentage of the offspring in order to spread the optimum solution search and avoid local optimal solutions.
\end{enumerate}

The computation is stopped when the population has converged to the same fitness value which is supposed to be the optimal solution. 

In the case of APSK the chromosomes are the radii and the phase of each symbol. The optimization criterion chosen is the minimization of the MSE, i.e. the expected minimum squared error between the transmitted symbol and the received one. Therefore the function used to calculate the fitness scores is $R=1/(MSE)$.
In \cite{angioi2010} the importance of a constellation design that takes into account UEP has been highlighted and GA has been demonstrated to be a viable solution for solving the OP. However the choice of the best selection, crossover and mutation functions together with an appropriate number of generation iterations and population size has not been exploited yet. For this reason, this paper extends the results obtained in \cite{angioi2010} demonstrating that a proper selection of the abovementioned parameters can further improve the optimality of the solution. In particular, this work concentrates on the best choice regarding the selection and the crossover function by comparing the results obtained through simulation in Matlab for 5 different selection functions (\textit{stochastic uniform, remainder, uniform, roulette, tournament}) and 6 different crossover functions (\textit{scattered, single point, two point, intermediate, heuristic, arithmetic}). For more information about the abovementioned functions, the reader can refer to the Matlab guide and a vast literature on the topic.

\section{Results for 16-APSK}

First of all, let us analyze the case of 16-APSK. The GA starts with an initial population of $80$ chromosomes characterized by genes with values uniformly distributed on the constellation circles. Using these values, transmission over satellite is simulated according to the model presented in Section II. Then, the fitness function is evaluated thanks to the function $R$ already discussed. At this point, the GA modifies the population in accordance with the fitness results using the policies defined by the specific selection and crossover functions taken into consideration. The transmission is then repeated using the new generation until convergence or the maximum number of generations (in this case set to $n=130$) is reached.\\

\begin{table}[h!]
\centering
\begin{tabular}{|c||c|c|c|c|c|}
    \hline
    \backslashbox{CR}{SEL}&stochunif&remainder&uniform&roulette&tournam.\\
    \hline \hline
    scattered&\textbf{1.1045}&1.2225&1.1724&1.2054&1.2287\\ 
    \hline
    single point&1.1758&\textbf{1.1029}&1.2541&\textbf{1.1111}&1.1577\\ 
    \hline
    two point&1.1336&1.1235&1.2720&1.2002&1.1470\\
    \hline
    intermediate&1.2074&1.1983&1.6869&1.1981&1.2149\\ 
    \hline
    heuristic&1.1695&1.1701&1.1706&1.1609&1.1514\\ 
    \hline
    arithmetic&1.2749&1.2968&1.8615&1.3074&1.1830\\ 
    \hline
  \end{tabular}
  \vspace{0.1cm}

\caption{Target values for 16-APSK 4sim}
\label{tab:16APSK4}
\end{table}

\begin{table}[h!]
\centering
\begin{tabular}{|c||c|c|c|c|c|}
    \hline
    \backslashbox{CR}{SEL}& stochunif & remainder & uniform & roulette & tournam. \\
    \hline \hline
    scattered &    1.2640  &  0.7629  &  1.0275  &  0.7439  &  0.7619\\ 
    \hline
    single point &    1.4284 &   \textbf{0.7259} &   0.9252  &  \textbf{0.7238}  &  0.9086\\ 
    \hline
    two point &    \textbf{0.7361} &   0.7527  &  0.7639 &   0.7942  &  0.7809\\
    \hline
    intermediate &    0.7732  &  0.8237  &  2.1824  &  0.7921  &  0.7530\\ 
    \hline
    heuristic &   0.7969  &  1.4852  &  0.8238  &  0.7967  &  0.7722\\ 
    \hline
    arithmetic &   0.7437  &  0.8052  &  3.3424  &  0.8634  &  0.7616\\ 
    \hline
  \end{tabular}
  \vspace{0.1cm}

\caption{Target values for 16-APSK 2sim}
\label{tab:16APSK2}
\end{table}

\begin{table}[h!]
\centering
\begin{tabular}{|c||c|c|c|c|c|}
    \hline
    \backslashbox{CR}{SEL}& stochunif & remainder & uniform & roulette & tournam. \\
    \hline \hline
    scattered &    0.9844  &  0.7686 &   2.3634 &   1.1534 &   1.1600\\ 
    \hline
    single point &0.8674 &   \textbf{0.6696}&    1.6173 &   0.8807  &  0.7660\\ 
    \hline
    two point & \textbf{0.7127} &   1.0152  &  3.0634  &  0.7291 &   0.7899\\
    \hline
    intermediate &0.8709  &  1.2898 &  15.5712 &   0.9861 &   1.5404\\ 
    \hline
    heuristic &   1.9504   & 1.3986 &   \textbf{0.6794}  &  0.7527  &  2.7479\\ 
    \hline
    arithmetic &     0.9338 &   1.0234&   13.9503  &  0.8291 &   1.1862\\ 
    \hline
  \end{tabular}
  \vspace{0.1cm}

\caption{Target values for 16-APSK 0sim}
\label{tab:16APSK0}
\end{table}

\begin{figure}[htbp!]
\centering
\includegraphics [width=1 \columnwidth] {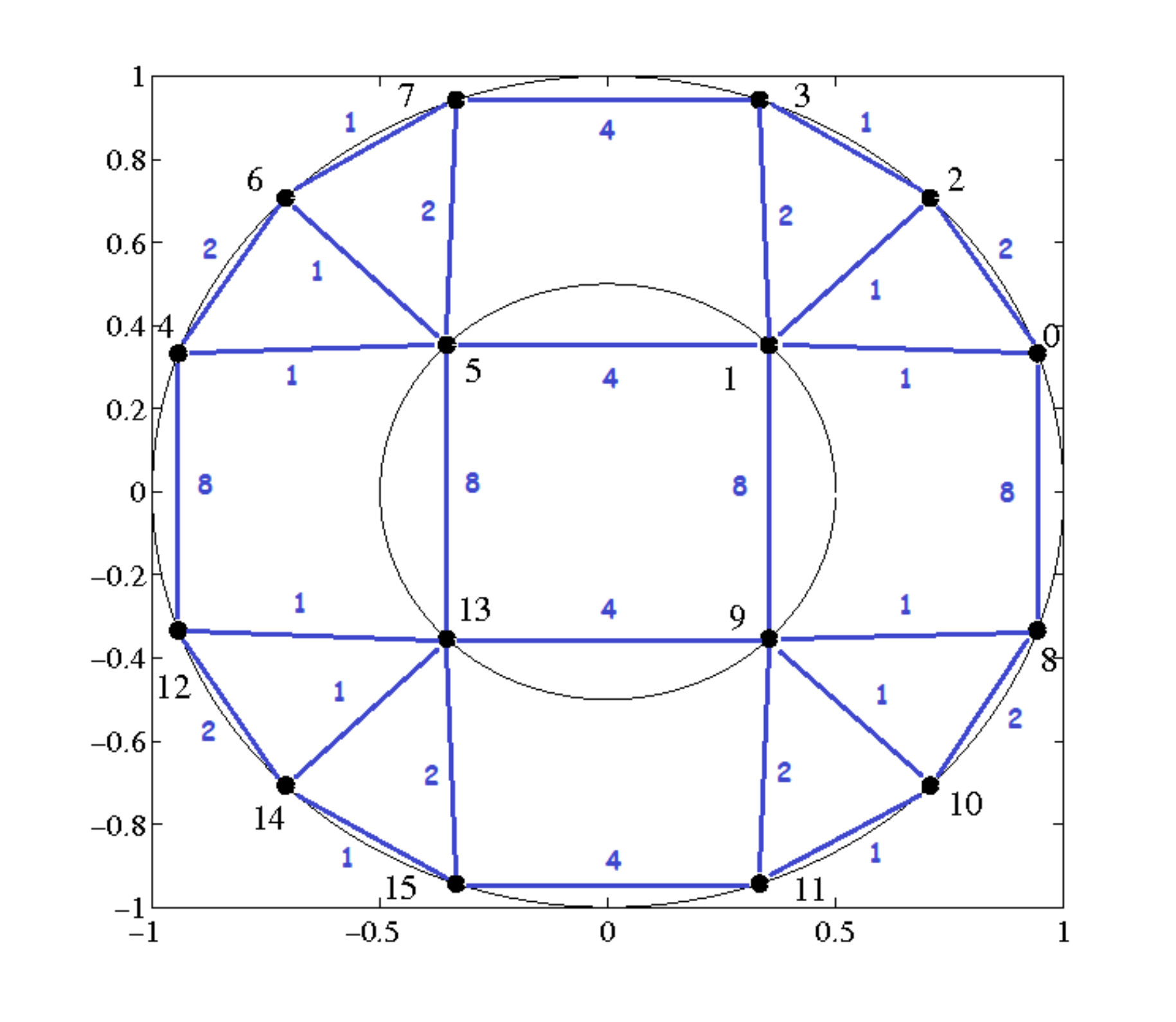}
\caption{Example of 16-APSK constellation with double symmetry}
\label{Ex4sim}
\end{figure}

In Tables \ref{tab:16APSK4},  \ref{tab:16APSK2} and  \ref{tab:16APSK0} the results in terms of $MSE$ at the target $SNR=10dB$ are shown respectively for the case of double symmetry ($x$ and $y$ axis), for the case with symmetry on only one axis and for the case without any symmetry constraint. The mentioned symmetry refers to the placement of the symbols with regard to the inter-symbol distorsion, as shown in Figure \ref{Ex4sim} for the case of double symmetry.

The best three results are highlighted in bold. It can be seen that regardless of the considered symmetries, the combination of the \textit{remainder} selection function and the \textit{single point} crossover function always yields to the best result or to a result really close to the best one. Moreover, from a general perspective, it can be gathered from the tables that the optimization procedure benefits from the drop of the symmetry contraints. As a matter of fact, comparing the case of double symmetry and the one without symmetry, the $MSE$ value at the target SNR is reduced by approximately $40\%$. However, in order to validate the results, it is important to verify whether the improvement extends for a certain $SNR$ range or if it is only local and specific of that $SNR$.

\begin{table}[tbh!]
\centering
\begin{tabular}{|c||c|c|c|c|}
    \hline
    \textbf{Parameter} & \textbf{[8]} & \textbf{double sym.} & \textbf{single sym.} & \textbf{no sym.} \\
    \hline \hline
    \textbf{$\rho_0$} &    0.6404  &  0.8996 & 0.9627 & 0.9593 \\ 
    \hline
    \textbf{$\theta_0$} & 1.0482 & 1.0360 & 2.5650 & 5.0872 \\ 
    \hline
    \textbf{$\theta_1$} &  0.9355 &    0.8867 & 2.3592 & 4.7453 \\ 
    \hline
    \textbf{$\theta_2$} & 0.6374 & 0.5802 & 2.0128 & 4.3400 \\
    \hline
    \textbf{$\theta_3$} & 0.4891  & 0.4013 & 1.7317 & 3.7447 \\ 
    \hline
    \textbf{$\theta_4$} & - & - & 1.4188  & 3.4121 \\ 
    \hline
    \textbf{$\theta_5$} & - & - & 1.2107 & 3.1109 \\ 
    \hline
    \textbf{$\theta_6$} & - & - & 0.8849 & 2.7071 \\ 
    \hline
    \textbf{$\theta_7$} & - & - & 0.5372 & 2.2326 \\ 
    \hline
    \textbf{$\theta_8$} & - & - & - & 1.8925 \\ 
    \hline
    \textbf{$\theta_9$} & - & - & - & 1.5490 \\ 
    \hline
    \textbf{$\theta_{10}$} & - & - & - & 1.2567 \\ 
    \hline
    \textbf{$\theta_{11}$} & - & - & - & 1.0438 \\ 
    \hline
    \textbf{$\theta_{12}$} & - & - & - & 0.7340 \\ 
    \hline
    \textbf{$\theta_{13}$} & - & - & - & 0.4687 \\ 
    \hline
    \textbf{$\theta_{14}$} & - & - & - & 0.2205 \\ 
    \hline
    \textbf{$\theta_{15}$} & - & - & - & 0.0699 \\ 
    \hline
  \end{tabular}
  \vspace{0.1cm}
\caption{Parameter values for 16-APSK}
\label{tab:values16}
\end{table}

For this reason, in Table \ref{tab:values16} the values for the radius and the phase of the various symbols are presented. The first column refers to the results obtained in \cite{angioi2010}. The second, third and fourth column refer respectively to the cases of double, single and no symmetry for the case in which the \textit{remainder} selection function and the \textit{single point} crossover function are used. The considered chromosomes for these three cases are respectively:\\

\begin{itemize}
\item $\gamma=[\rho_0,\theta_0,\theta_1,\theta_2,\theta_3]$\\
\item $\gamma=[\rho_0,\theta_0,\theta_1,\theta_2,\theta_3,\theta_4,\theta_5,\theta_6,\theta_7]$\\
\item $\gamma=[\rho_0,\theta_0,\theta_1,\theta_2,\theta_3,\theta_4,\theta_5,\theta_6,\theta_7,\theta_8,\theta_9,\theta_{10},\theta_{11},\theta_{12},\theta_{13},\\
\theta_{14},\theta_{15}]$\\
\end{itemize}
where the different number of parameters is due to the fact that, when using symmetries, the rest of the symbols are defined by symmetry. Notice also that only one radius has been defined in the table, since we are assuming that the outer one is fixed to $1$. In addition, the phases have been defined so that the subscripts of each theta correspond to the alphabet value assigned to that symbol.

\begin{figure}[htbp!]
\centering
\includegraphics [width=1 \columnwidth] {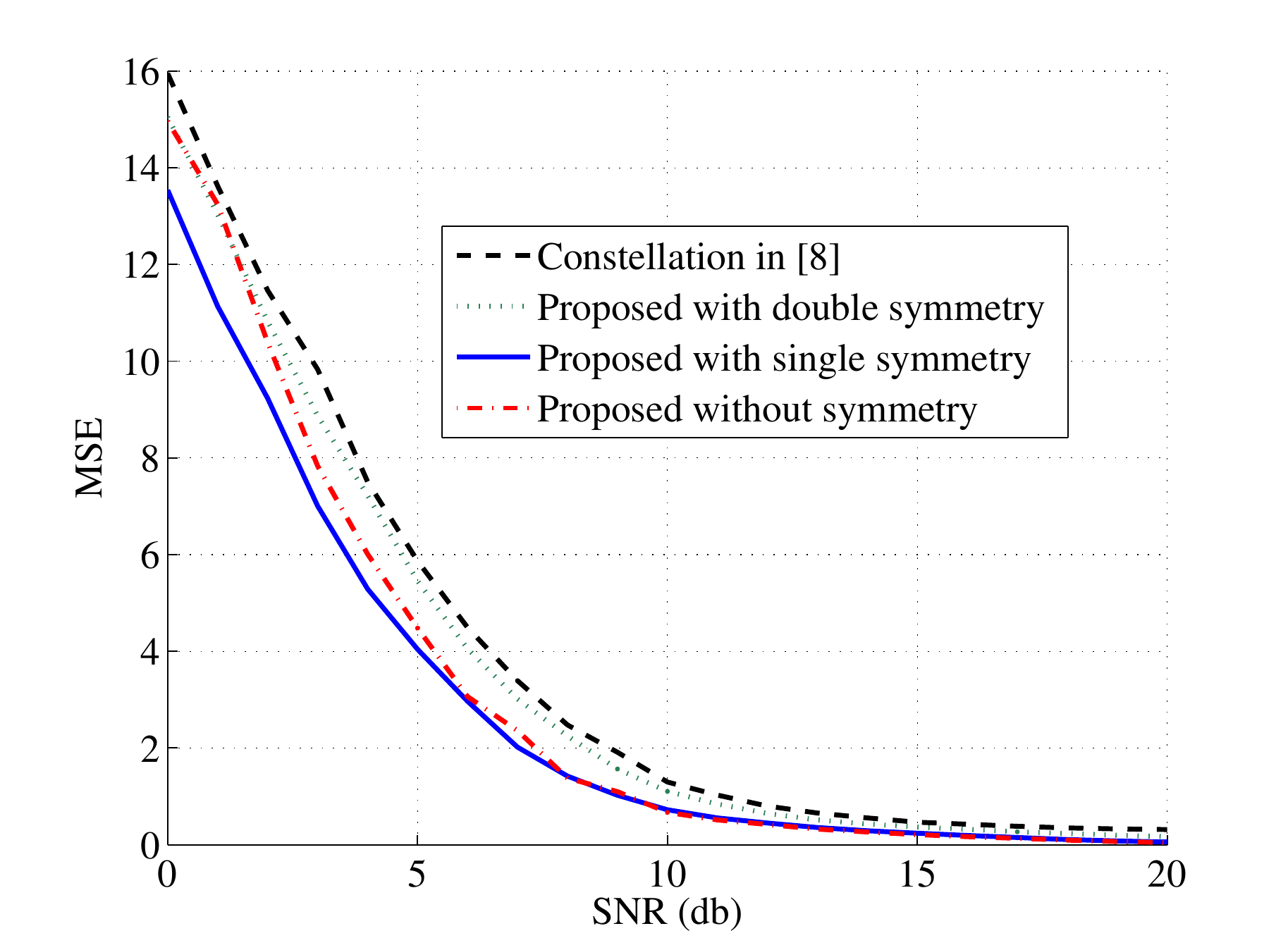}
\caption{MSE results for the considered constellations}
\label{16comp}
\end{figure}

\begin{figure}[htbp!]
\centering
\includegraphics [width=1 \columnwidth] {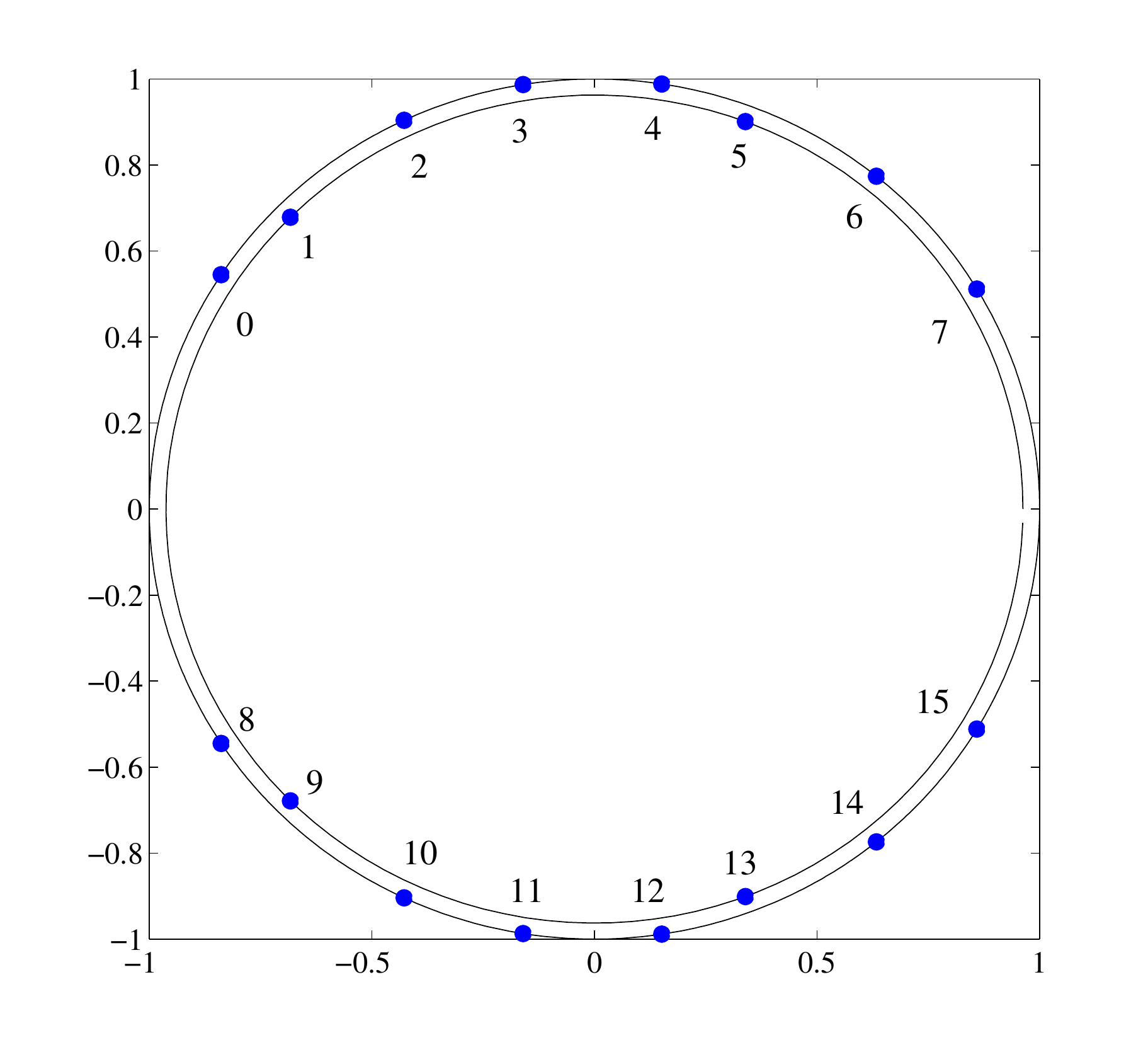}
\caption{Optimized constellation for 16-APSK with single symmetry}
\label{16cost_2sim}
\end{figure}

Figure \ref{16comp} shows the results in terms of $MSE$ as a function of the $SNR$ for the 4 constellations presented in Table \ref{tab:values16}. Unfortunately, in \cite{angioi2010} it was not stated what kind of selection and crossover functions were used. However, it can be seen from the graph that a proper choice of the functions results in a better performance even when both symmetries are kept. Surprisingly, the results for a single symmetry are better than those with no symmetries, although this last case overtakes the first one from $SNR=10dB$. These results demonstrate that having a better $MSE$ at the $SNR$ target and/or dropping all the symmetry constraints does not necessarily corresponds to an improvement of the performance. In figure \ref{16cost_2sim}, the constellation with single symmetry is shown.

\section{Results for 32-APSK}

The same procedure presented in the previous section has also been applied to the 32-APSK constellation, with the necessary modifications. The first difference that can be noticed is the greater number of variables to compute. Therefore, also the convergence of the optimization is expected to be slower than in the previous case. Nevertheless, we decided to keep the population size to the value $80$ and the number of generations to $130$ in order to evaluate how optimal the solutions are keeping the same values as in 16-APSK. Moreover, this has been dictated by time constraints, since in the case of 32-APSK with no symmetries each combination of selection and crossover function required approximately 8 hours to be run.

\begin{table}[tbh!]
\centering
\begin{tabular}{|c||c|c|c|c|c|}
    \hline
    \backslashbox{CR}{SEL}& stochunif & remainder & uniform & roulette & tournam. \\
    \hline \hline
    scattered &4.5572 &   4.6612  &  5.2848  & \textbf{4.4263} &  4.6581\\ 
    \hline
    single point &4.8225 &   4.9691  &  5.3944  &  4.5561  &  4.5573\\ 
    \hline
    two point &4.7172   & 4.7339  &  5.3761  &  4.7521  &  \textbf{4.4136}\\
    \hline
    intermediate &5.2306  &  5.4009  & 13.1101  &  5.5408  &  4.7368\\ 
    \hline
    heuristic &4.7713  &  4.5942   & 4.5201  &  5.2484  &  4.8416\\ 
    \hline
    arithmetic &5.8264  &  5.4936  & 12.0061  &  5.4959  &  \textbf{4.4832}\\ 
    \hline
  \end{tabular}
  \vspace{0.1cm}

\caption{Target values for 32-APSK 4sim}
\label{tab:32APSK4}
\end{table}

\begin{table}[tbh!]
\centering
\begin{tabular}{|c||c|c|c|c|c|}
    \hline
    \backslashbox{CR}{SEL}& stochunif & remainder & uniform & roulette & tournam. \\
    \hline \hline
    scattered &8.1330 &   4.8600  & 12.7564  &  5.1381   &  \textbf{4.3409}\\ 
    \hline
    single point &9.2198 &   5.5814  & 10.9982  & 10.9608  &  \textbf{4.1631}\\ 
    \hline
    two point &4.5536   & 4.4141 &  10.2918  &  4.4285  & \textbf{4.0857}\\
    \hline
    intermediate &4.6031  &  8.6994  & 11.7529 &  11.0164  &  4.9557\\ 
    \hline
    heuristic &4.3239  &  4.5903  &  5.1924  &  6.6715  &  4.78346\\ 
    \hline
    arithmetic &5.1618  &  5.2917  & 13.1309 &  10.7594 &   4.6443\\ 
    \hline
  \end{tabular}
  \vspace{0.1cm}
\caption{Target values for 32-APSK 2sim}
\label{tab:32APSK2}
\end{table}

\begin{table}[tbh!]
\centering
\begin{tabular}{|c||c|c|c|c|c|}
    \hline
    \backslashbox{CR}{SEL}& stochunif & remainder & uniform & roulette & tournam. \\
    \hline \hline
    scattered &6.3820  &  7.4839 &  25.7629  &  7.8117  &  \textbf{5.9957}\\ 
    \hline
    single point &8.7097  & 14.4059 &  26.9502  &  6.3373  &  6.8377\\ 
    \hline
    two point &15.6945   & 7.0951 &  19.4484  & \textbf{5.6768}  & \textbf{5.6494}\\
    \hline
    intermediate &12.9701 &  23.4311 &  72.1346  & 17.7114  & 14.7770\\ 
    \hline
    heuristic &12.1008  &   9.6265  &  10.5484 &  26.8005  &  9.5540\\ 
    \hline
    arithmetic &43.5802  & 36.4328  & 42.8421  & 11.6793  &  8.1312\\ 
    \hline
  \end{tabular}
  \vspace{0.1cm}
\caption{Target values for 32-APSK 0sim}
\label{tab:32APSK0}
\end{table}

Tables \ref{tab:32APSK4}, \ref{tab:32APSK2} and \ref{tab:32APSK0} show that, not unexpectedly, when the number of variables to compute increases the algorithm is not anymore able to converge to a solution in the number of generations set. Moreover, each combination of the selection and crossover functions converges with different paces. Another interesting result is that, when the number of symbols is changed, the best result is not obtained for the same selection and crossover function. We expect the same thing to hold when some symbols are moved from one radius to another. The proof of this expectation and a deeper explanation of why this happens are left as future research on the topic.

\begin{table}[htbp!]
\centering
\begin{tabular}{|c||c|c|c|}
    \hline
    \textbf{Parameter} & \textbf{[8]} & \textbf{double sym.} & \textbf{single sym.}\\
    \hline \hline
    \textbf{$\rho_0$} &    0.2453  &   0.2446 &   0.2487\\ 
    \hline
    \textbf{$\rho_1$} &    0.8163  &  0.8285 & 0.8217\\ 
    \hline
    \textbf{$\theta_0$} & 3.9215 & 0.1664 &   0.5301\\
    \hline
    \textbf{$\theta_1$} & 3.8878 &  0.2998 & 0.7360\\ 
    \hline
    \textbf{$\theta_2$} & 3.7697 & 0.3009 &  0.7342\\
    \hline
    \textbf{$\theta_3$} & 3.6837  & 0.6293 & 0.8993\\ 
    \hline
    \textbf{$\theta_4$} & 3.4184 & 0.5831 &  1.1148\\ 
    \hline
    \textbf{$\theta_5$} & 3.6422 & 0.9550 & 1.2771\\ 
    \hline
    \textbf{$\theta_6$} & 3.2628 & 0.9219 & 1.4283\\ 
    \hline
    \textbf{$\theta_7$} & 3.1881 & 1.0028 & 1.5063\\ 
    \hline
    \textbf{$\theta_8$} & 2.6639 & - & 1.8236\\ 
    \hline
    \textbf{$\theta_9$} & 2.6409 & - & 1.9895\\ 
    \hline
    \textbf{$\theta_{10}$} & 2.4866 & - & 2.0694\\ 
    \hline
    \textbf{$\theta_{11}$} & 2.3709 & - & 2.1592\\ 
    \hline
    \textbf{$\theta_{12}$} & 2.2034 & - & 2.2590\\ 
    \hline
    \textbf{$\theta_{13}$} & 2.1479 & - & 2.5016\\ 
    \hline
    \textbf{$\theta_{14}$} & 2.0492 & - & 2.5479\\ 
    \hline
    \textbf{$\theta_{15}$} & 2.0199 & - & 2.5813\\ 
    \hline
  \end{tabular}
  \vspace{0.1cm}
\caption{Parameter values for 32-APSK}
\label{tab:values32}
\end{table}

In Table \ref{tab:values32} the values for the parameters describing the position of the constellation symbols are given. In this case, the case without symmetry has not been included due to the lack of significance. Concerning the differences between the angle ranges of our case and the one in \cite{angioi2010}, this is simply due to the fact that we have considered $[0,\pi]$ as our optimization range while in \cite{angioi2010} the considered range was $[\pi/2,3\pi/2]$. 

\begin{figure}[htbp!]
\centering
\includegraphics [width=1 \columnwidth] {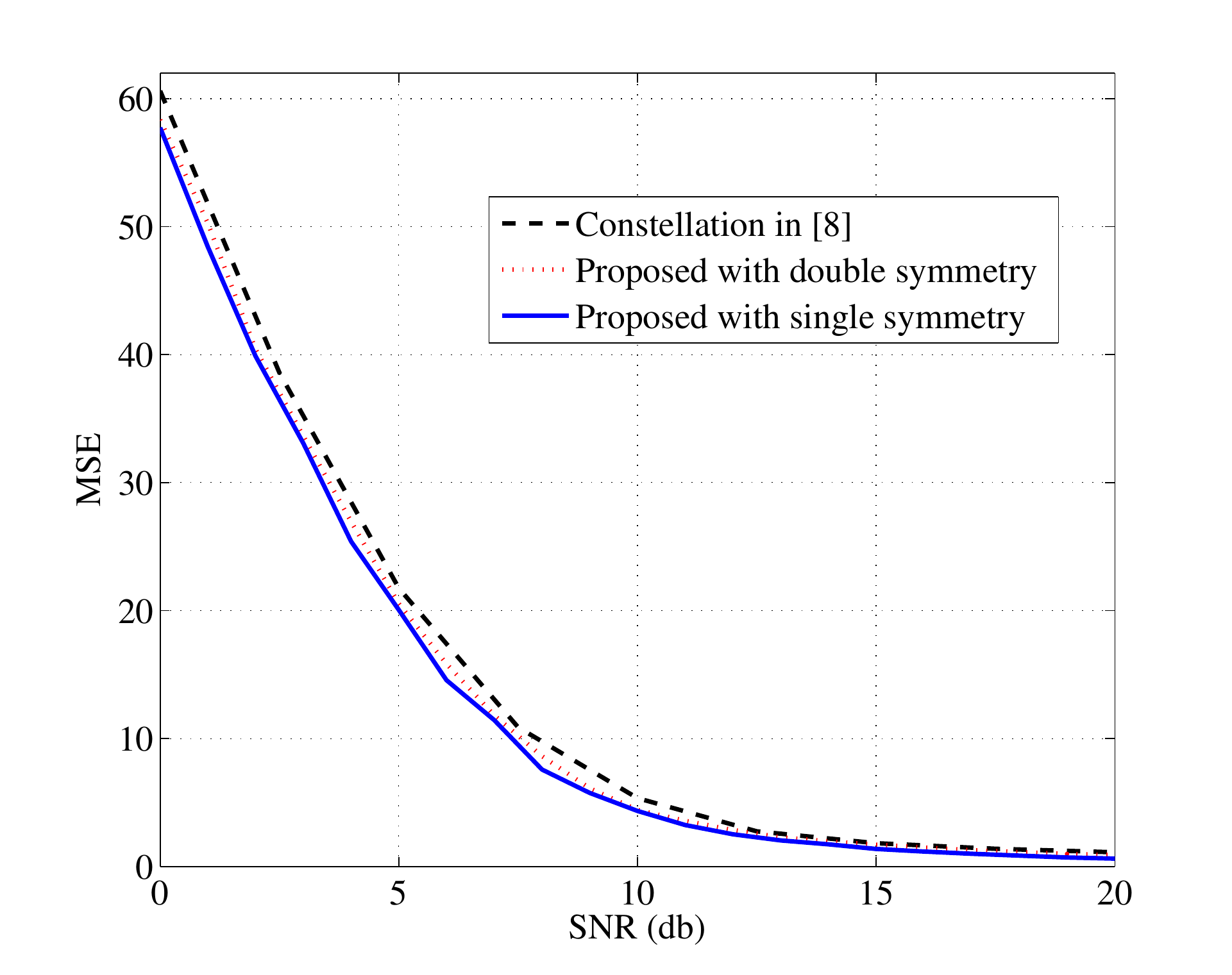}
\caption{MSE results for the considered constellations}
\label{32comp}
\end{figure}

\begin{figure}[htbp!]
\centering
\includegraphics [width=1 \columnwidth] {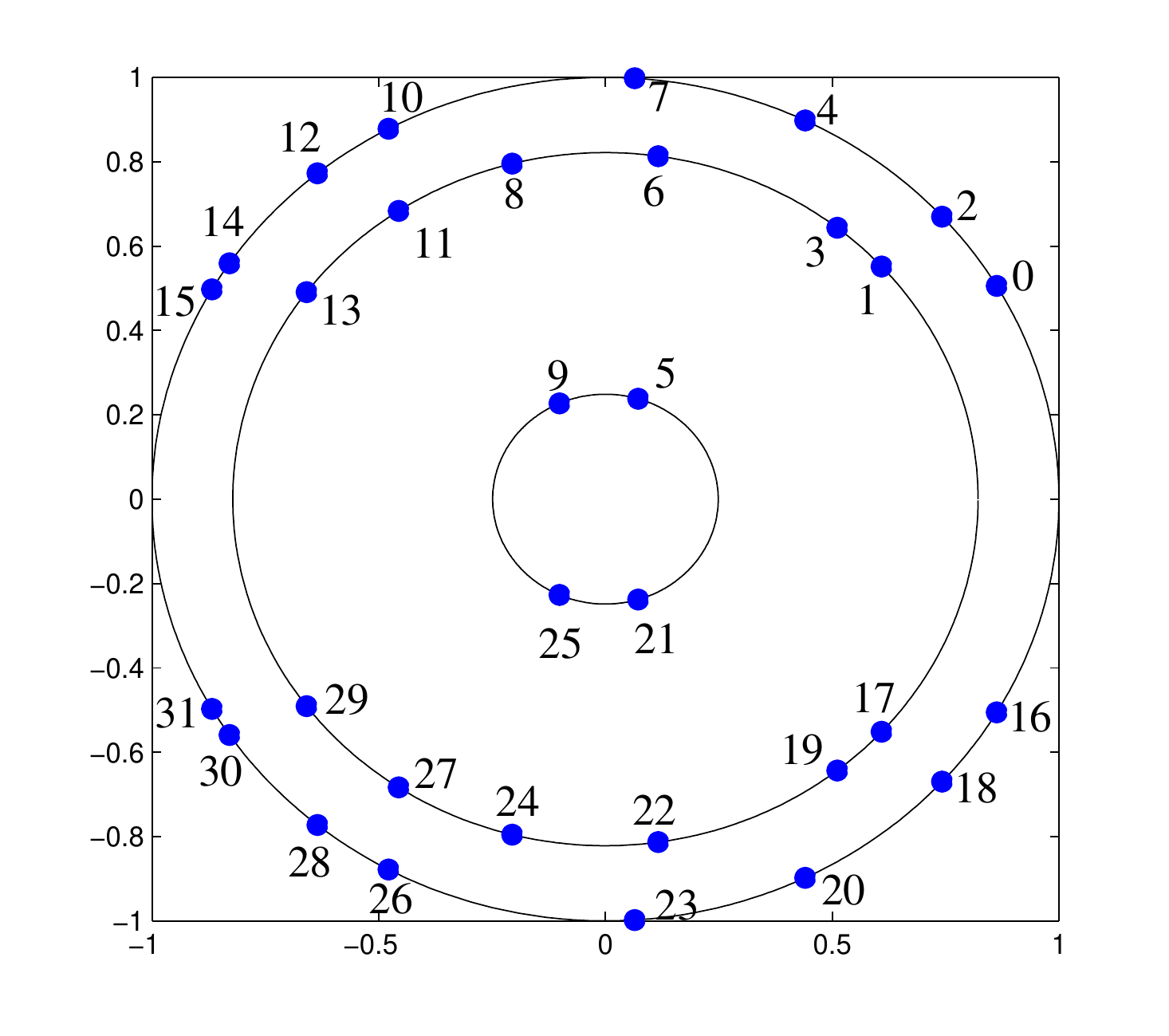}
\caption{Optimized constellation for 16-APSK with single symmetry}
\label{32cost_2sim}
\end{figure}

Figure \ref{32comp} presents the results for the constellations described in Table \ref{tab:values32}. Although in a smoother way, also in this case the obtained results improve those obtained in \cite{angioi2010}. In particular, the $MSE$ is lowered from 5.34 to 4.34, that is approximately a decrement of the 19\%. Figure \ref{32cost_2sim} shows the proposed constellation for 32-APSK with single symmetry.

\section{Conclusions and Future Work}

In this paper, the use of asymmetric constellations for uncoded transmission in the satellite broadcasting scenario has been presented. Moreover, their optimization using genetic algorithms and the careful selection of the functions involved in the optimization routine have been discussed. Found results demonstrate that it is possible to further optimize this kind of communications by adjusting the behavior of the genetic algorithm. Moreover it has been demonstrated that dropping symmetry constraints is not always beneficial to the optimization process, especially when several variables must be computed thus slowing down the convergence to an optimal solution of the genetic algorithm. As future work, we aim at extending the results presented in this paper as well as apply the same concepts to the case of 64-APSK.

\end{document}